\documentstyle[prl,aps,twocolumn,psfig]{revtex}

 \newcommand{\CL}{{\cal L}}
 
 \newcommand{\CN}{{\cal N}}
 \newcommand{\CO}{{\cal O}}

 \newcommand{\bea}{\begin{eqnarray}}  \newcommand{\eea}{\end{eqnarray}}
 \newcommand{\beq}{\begin{equation}}  \newcommand{\eeq}{\end{equation}}
 \newcommand{\non}{\nonumber}  
   
 \newcommand{\lmk}{\left(}  \newcommand{\rmk}{\right)}
 \newcommand{\lkk}{\left[}  \newcommand{\rkk}{\right]}
   
 \newcommand{\del}{\partial}  
 \newcommand{\vect}[1]{\mbox{\boldmath${#1}$}}
 \newcommand{\bib}{\bibitem}

 \newcommand{\gtilde} {~ \raisebox{-1ex}{$\stackrel{\textstyle >}{\sim}$} ~} 
 \newcommand{\ltilde} {~ \raisebox{-1ex}{$\stackrel{\textstyle <}{\sim}$} ~}

\begin{document}
\twocolumn[\hsize\textwidth\columnwidth\hsize\csname
@twocolumnfalse\endcsname
\draft
\title{Evolution of Axionic Strings and Spectrum of Axions Radiated from Them}
\author{Masahide Yamaguchi$^{1}$, M. Kawasaki$^{2}$\footnotemark[1],
  and Jun'ichi Yokoyama$^{3}$\footnotemark[2]}
\address{$^{1}${\it Department of Physics, University of Tokyo, Tokyo~113-0033, Japan}}
\address{$^{2}${\it Institute for Cosmic Ray Research, University of
    Tokyo, Tanashi~188-8502, Japan}}
\address{$^{3}${\it Yukawa Institute for Theoretical Physics, Kyoto
    University, Kyoto~606-8502, Japan}} 
\date{\today}
\maketitle

\begin{abstract}

Cosmological evolution of axionic strings is investigated by
numerically solving field equations of a complex scalar field in 3+1 
dimensions. It is shown that the global strings relax to the scaling
solution with a significantly smaller number density than the case of
local strings. The power spectrum of axions radiated from them is
calculated from the simulation data, which is found to be highly
peaked around the Hubble scale, and a more accurate constraint on the
Peccei-Quinn breaking scale is obtained.

\end{abstract}

]

\footnotetext[1]{Present address : \it{Research Center for the Early
    Universe, University of Tokyo, Tokyo~113-0033, Japan}}
\footnotetext[2]{Present address : {\it Department of Earth and Space
    Science, Graduate School of Science, Osaka Universe, Toyonaka,
    Osaka~560-0043, Japan}}

The global Peccei-Quinn(PQ) $U(1)$-symmetry is introduced into the
standard model of particle physics in order to solve the strong CP
problem of the quantum chromodynamics (QCD) \cite{PQ}. The axion
appears as a Nambu-Goldstone boson in consequence of the spontaneous
breakdown of such a global symmetry, and acquires a mass through the
instanton effect at the QCD scale \cite{WW}.

This breaking scale, $f_{a}$, is constrained by the terrestrial and
astrophysical experiments as well as by cosmological considerations.
The most stringent lower bound has been obtained from SN 1987A as
$f_{a} \gtilde 10^{10}$GeV \cite{SN}. On the other hand, the upper
bound is given by the condition that the present energy density due to the
axion should not overclose the universe.

In the very early universe, the $U(1)_{{\rm PQ}}$-symmetry is expected
to be restored due to the high temperature effects. As the universe
cools down, it spontaneously breaks down and global strings\,(axionic
strings) are formed \cite{VE}. Subsequent evolution of global strings
has been less examined than that of local strings, and the result for
the latter has been applied to axionic strings without direct
numerical verification. In particular, global strings have been
assumed to obey a scaling solution like local ones, in which the
energy density of infinite strings is given by

\beq
   \rho = \xi \mu / t^2,
\eeq
where $\mu$ is the average energy per unit length of strings and $\xi$
is a constant. It has been found that $\xi \sim 13$ for local strings in
the radiation dominant era \cite{BBAS}. 

But there is a crucial difference between global and local strings;
that is, the former strongly couples with the associated
Nambu-Goldstone fields. As a result, global strings have the following
prominent features: (a)\,These fields carry most of the energy rather
than string cores, (b)\,long-range forces proportional to the inverse
separation work between strings, and (c)\,the dominant energy-loss
mechanism of loops is radiation of Nambu-Goldstone bosons \cite{GB}.
Therefore it is not a trivial problem whether the global string
network obeys the scaling solution like local ones. In fact, a
deviation from scaling property is observed in 2+1 dimensions
\cite{2dim,YYK}, where ``strings'' cannot intercommute but only
pair-annihilate. Of course, since strings intercommute with the
probability of order unity \cite{SHE} in 3+1 dimensions, the results
in 2+1 dimensions do not apply directly.  In investigating the
property of a global string, the effective action is often used, which
is valid when massive modes are sufficiently damped. For a global
string we have to use the Kalb-Ramond action \cite{KR}, which is much
harder to deal with than the Nambu-Goto action for a local string.
Hence in the present Letter, instead of solving equations of motion
derived from the Kalb-Ramond action, we numerically solve the
evolution equation of the complex scalar field to trace the evolution
of the global string network and calculate the spectrum of axions
radiated from them. (If the reheating temperature after inflation is
lower than the symmetry breaking scale, global strings and radiated
axions are washed out and relic axions come only from coherent
oscillations of the zero mode \cite{zero}. Here we assume that this is
not the case.)

Axions are massless when they are emitted from strings and hence they
behave like radiation. Later around the QCD scale, they acquire a
small mass, $m_{a}$, through QCD instanton effects. Since at present
kinetic energy of axions is much smaller than the rest mass, the
present energy density of axions, $\rho_{a}$, is given by $\rho_{a} =
m_{a}n_{a}$ with $n_{a}$ being their number density. Thus, in order to
estimate the present energy density of axions, we must transform the
energy density of radiated axions into the number density by use of
the spectrum.  However, the spectral shape is still in dispute.
Davis, Shellard, and co-workers insist that the spectrum has a sharp
peak at the Hubble horizon scale \cite{DA,sharp,BS}. On the other
hand, Sikivie, and co-workers claim that it is proportional to the
inverse momentum \cite{flat}.  We clarify the shape of this spectrum
in realistic situations where global strings evolve according to the
scaling solution; that is, we identify kinetic energy of axions
emitted from strings and elucidate the spectrum by Fourier
transforming them. Hence our approach is free from the ambiguities
associated with the ad hoc choice of the initial configuration of
strings.

We consider the following Lagrangian density for a complex scalar
field $\Phi(x)$,
\beq
  \CL[\Phi] = \del_{\mu}\Phi \del^{\mu}\Phi^{\dagger} - V_{\rm eff}[\Phi] \:.
\eeq

\noindent
with the potential $V_{\rm eff}[\Phi]$ given by
\beq
  V_{\rm eff}[\Phi] = \frac{\lambda}{2}(\Phi\Phi^{\dagger} - \eta^2)^2 
                 + \frac{\lambda}{3}T^2\,\Phi\Phi^{\dagger} \:.
\eeq

\noindent
Hereafter we set $\lambda = 1.0$ for brevity. For $T > T_{c} =
\sqrt{3}\eta$, the potential $V_{\rm eff}$ has a minimum at $\Phi =
0$, and the $U(1)$ symmetry is restored. On the other hand, new minima
$|\Phi|_{{\rm min}} = \eta\sqrt{1-(T/T_{c})^2}$ appear and the
symmetry is broken for $T < T_{c}$. In this case the phase transition
is of second order.

In the flat Friedmann universe, the equation of motion is given by
\beq
  \ddot{\Phi}(x) + 3H\dot{\Phi}(x) - \frac{1}{a(t)^2}\nabla^2\Phi(x)
   = - V'_{\rm eff}[\Phi] \:,
\eeq

\noindent
where the prime represents the derivative $\del/\del\Phi^{\dagger}$
and $a(t)$ is the scale factor. The Hubble parameter $H = \dot
a(t)/a(t)$ and the cosmic time $t$ are given by
\bea
  H^2 = \frac{8\pi}{3 M_{\rm Pl}^2} \frac{\pi^2}{30} g_{*} T^4 \:,
   ~~~~~
  t = \frac{1}{2H}  \:,
\eea

\noindent
where $M_{\rm Pl}$ is the Planck mass, $g_{*}$ is the total effective
number of relativistic degrees of freedom, and radiation domination is
assumed. We define the dimensionless parameter $\zeta$ as 
\beq 
\zeta\equiv 
\lmk \frac{45M_{\rm Pl}^2}
{16\pi^3g_{*} \eta^2} \rmk^{1/2} \:.  
\eeq

\noindent  
In our simulation, we take $\zeta = 10$ and $g_{*} = 1000$, which
corresponds to $\eta \sim 10^{16}$ GeV, but the essential result is
independent of this choice.

We take the initial time $t_{i} = t_{c}/4$ and the final time $t_{f} =
75\,t_{i} = 18.75\,t_{c}$, where $t_c$ is the epoch $T=T_c$. 
Since the $U(1)$ symmetry is restored at
the initial time $t = t_{i}$, we adopt as the initial condition a
thermal equilibrium state with $\Phi$'s mass equal to the inverse
curvature of the potential at the origin. The scale factor $a(t)$ is
normalized as $a(t_{i}) = 1$.

\begin{figure}[htb]
  \begin{center}
    \leavevmode\psfig{figure=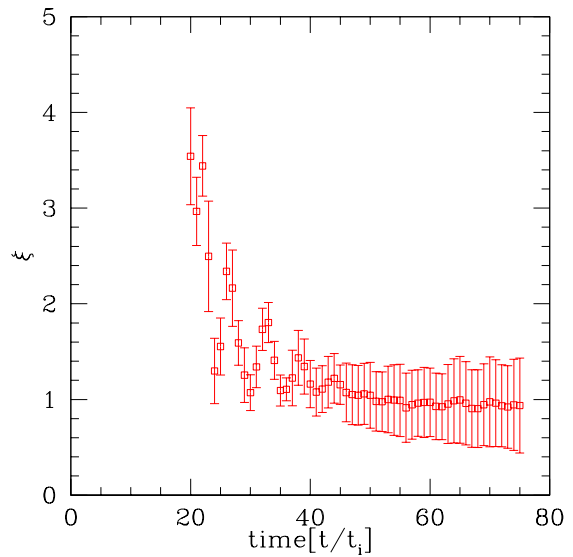,width=7.59cm}
  \end{center}
  \caption{Symbols($\Box$) represent time development of $\xi$. The vertical lines denote a standard deviation over ten different initial
  conditions.}
  \label{fig:statistics}
\end{figure}

We perform the simulations in $256^3$ lattices with the physical
lattice spacing $\delta x_{\rm phys} = \sqrt{3}t_{i} a(t)/25$. The
time step is taken as $\delta t = 0.01 t_{i}$. Hereafter the subscript
``phys'' is omitted.  The box size is nearly equal to the Hubble
horizon $H^{-1}$ and a typical width $d \sim 1.0/(\sqrt{2}\eta)$ of a
string is twice as large as the lattice spacing at the final time
$t_{f}$.  We simulate the system from 10 different thermal initial
conditions using the second order leap-frog method and the
Crank-Nicholson scheme. We use the same method to identify a string
core as in our previous work \cite{YYK}. More details will be
published elsewhere \cite{Yama}. The zero temperature potential
$V_{\rm eff}[\Phi, T = 0]$ is used after $t = 20 t_{i}$ so that the
axion is identified as in Eq.(\ref{phasedef}) below. We find that
after some relaxation period the energy density of strings $\rho$ is
given by

\beq
   \rho = \xi \mu / t^2,
\eeq

\noindent
where $\xi \simeq (1.00 \pm 0.08)$ irrespective of time and $\mu
\equiv 2\pi\eta^2\ln(t/(d\xi^{1/2}))$ is the average energy per unit
length of strings.  Therefore, we can conclude that global strings
network relax into scaling regime. In Fig.\ \ref{fig:statistics}, we
show time development of $\xi$.

Now we turn to the spectrum of axions emitted from axionic strings
under the situation where they follow the scaling solution. If we
represent the complex field $\Phi(t, \vect x)$ in terms of the radial
mode $\chi(t, \vect x)$ and the axion field $\alpha(t, \vect x)$ as

\beq
  \Phi(t,\vect x) = \lkk \eta + \frac{\chi(t,\vect x)}{\sqrt2} \rkk 
               \exp \lmk \frac{i\alpha(t, \vect x)}{\sqrt2 \eta} \rmk \:,
  \label{phasedef}
\eeq    

\noindent
the kinetic energy density of axions is given by

\bea
  \frac12 \dot\alpha(t, \vect x)^2 &=& 
      \frac{\eta^2}{|\Phi(t, \vect x)|^4} \times
   \non \\
   && \hspace{-1.5cm} 
     \lkk - {\rm Im}\Phi(t, \vect x) {\rm Re}\dot\Phi(t, \vect x)
          + {\rm Re}\Phi(t, \vect x) {\rm Im}\dot\Phi(t, \vect x) 
     \rkk^2 \:.
\eea 

Since emitted axions damp like radiation, the average energy density
of axions radiated in the period between $t_{1}$ and $t_{2}$,
$\bar\rho[t_{1},t_{2}]$, is given by

\bea
  \bar{\rho}[t_{1},t_{2}] &=& 
  \frac{1}{V} \int d^3\vect x \rho[t_{1},t_{2};\vect x] \non \\* 
                          &=& 
       \frac{1}{V} \int d^3\vect x \lkk 
           \frac12 \dot\alpha(t_{2}, \vect x)^2 -
           \frac12 \dot\alpha(t_{1}, \vect x)^2 
                                   \lmk
           \frac{t_{1}}{t_{2}} 
              \rmk^2 \rkk 
  \non \\
                            &=&
       \frac{1}{V} \int \frac{d^3\vect k}{(2\pi)^3} 
       \lkk \frac12\, |\dot\alpha_{\vect k}(t_{2})|^2 -
            \frac12\, |\dot\alpha_{\vect k}(t_{1})|^2 \lmk 
            \frac{t_{1}}{t_{2}} \rmk^2 \rkk 
  \non \\
                            &\equiv&
  \int \frac{d^3\vect k}{(2\pi)^3} 
         \tilde{\rho}_{\vect k}[t_{1},t_{2}]
                             \equiv
  \int_{0}^{\infty} \frac{dk}{2\pi^2} {\rho_{k}}[t_{1},t_{2}] 
\:,
\label{eqn:spectrum}
\eea

\noindent
where $V$ is the simulation volume and $\dot\alpha(t, \vect x) = \int
\frac{d^3\vect k}{(2\pi)^3} \dot{\alpha}_{\vect k}(t) \exp(i \vect k
\cdot \vect x)$.

\begin{figure}
  \begin{center}
    \leavevmode\psfig{figure=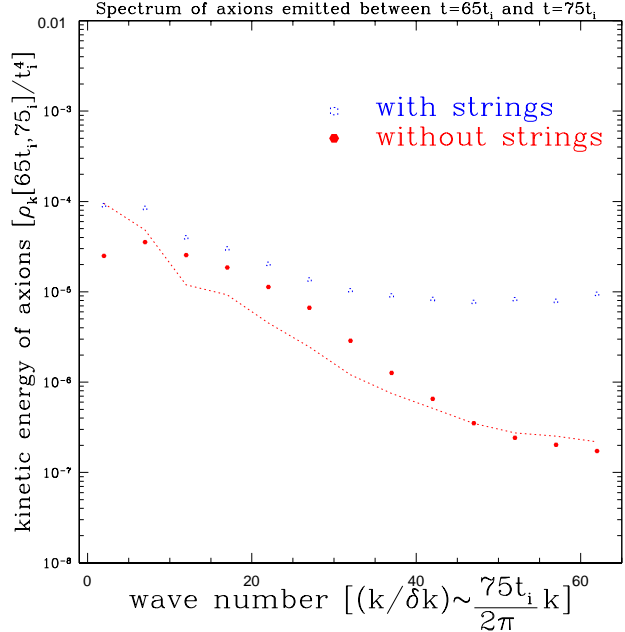,width=8.6cm}
  \end{center}
  \caption{Filled dots represent the power spectrum, $\rho_k$, which
    is already averaged over the direction of $\vect{k}$ and 
    multiplied by the phase-space factor as defined in
    Eq.(\ref{eqn:spectrum}). It is obtained from cells with no string
    cores. The dotted line denotes its standard deviation.  Open dots
    represent the spectrum obtained by averaging
     over all cells including string cores.
    Bins are cut every $5\delta k$. $k = 64\delta k$ corresponds to
    string cores.}
  \label{fig:spectrum}
\end{figure}

To avoid contamination of string cores to the spectrum of emitted
axions, we divide the simulation box into 8 cells and stock the field
data of a cell if there are no string cores in that cell between
$t_{1}$ and $t_{2}$. Over all such cells, we average power spectra of
kinetic energy of axions obtained through Fourier transformation.  We
follow the above procedure between $t_{1} = 65t_{i}$ and $t_{2} =
75t_{i}$, whose result is depicted in Fig.\ \ref{fig:spectrum}.

As is seen there, the spectrum, $\rho_{k}$, is peaked around the
horizon scale and decays exponentially for higher momenta. Note that
it is contributed from both infinite and loop strings.  For
comparison, we have also given the power spectrum averaged over all
cells, some of which include string cores. Note that effects of string
cores lift the spectrum at high momenta and make it flatter.

Defining the average inverse momentum by

\beq
  \overline{k^{-1}} = 
       \frac{\int_{0}^{\infty} \frac{dk}{2\pi^2} \frac{\rho_{k}}{k}}
            {\int_{0}^{\infty} \frac{dk}{2\pi^2} \rho_{k}}
  \:,
\eeq

\noindent
we find $\overline{k^{-1}} \simeq (0.25\pm0.18)\times(t/2\pi)$. 

Finally, we want to estimate energy density of relic axions radiated
from axionic strings. Though dynamic range of our simulation is
limited  to
 $\ln(t/d) \sim 5$, we extrapolate our results up to the
cosmological scale with $\ln(t/d) \sim 75$. From the scaling property,
we can estimate the energy density of axions radiated from strings
from $\tau = t_{*}$ to $\tau = t$,

\beq
  \rho_{a}(t) = 2\pi f_{a}^2 \frac{\xi}{t^{2}} \int_{t_{*}}^{t}
                \frac1{\tau} \lkk \ln \lmk \frac{\tau}{d\xi^{1/2}} \rmk
                                 - 1 \rkk d\tau \:,
\eeq

\noindent
where $t_{*}$ is the cosmic time when axionic strings began radiating
axions and we denote the breaking scale, $\eta$, by $f_{a}$.
Multiplying it by the average inverse momentum
$\overline{k^{-1}(\tau)} \equiv \tau/(2\pi\epsilon)$, the number
density of relic axions is given by

\bea
  n_{a}(t) &=& 2\pi f_{a}^2 \frac{\xi}{t^{2}} \int_{t_{*}}^{t}
                \frac1{\tau} \lkk \ln{\lmk\frac{\tau}{d\xi^{1/2}} 
                                          \rmk}
                                 - 1 \rkk \overline{k^{-1}(\tau)}
                d\tau \non \\
           &\sim& \frac{f_{a}^2}{t} \frac{\xi}{\epsilon} \ln
               \lmk \frac{t}{d\xi^{1/2}} \rmk \:.
  \label{eqn:number}
\eea

As the temperature cools down to the QCD scale and the axion acquires
a non-vanishing mass, a network of domain-walls bounded by strings is
created and walls start to dominate the dynamics of the system at $t =
t_{w}$ given by \cite{BS},

\bea
  t_{w} &\sim& 1.7 \times 10^{-6} \Delta^2 
         \lmk \frac{m_{a}}{6\times10^{-6} {\rm eV}}
             \rmk^{-2} \lmk \frac{f_{a}}{10^{12} {\rm GeV}}
              \rmk^{-1.64} \non \\
         && \times \lmk \frac{\CN_{\rm QCD}}{60} \rmk^{0.5} {\rm sec}
  \:.
\eea

\noindent
where $\Delta$ is a constant of order unity which describes
uncertainties at the QCD phase transition,

\bea
  \Delta &=& 10^{\pm0.5} \lmk \frac{m_{a}}{6\times10^{-6} {\rm eV}}
             \rmk^{0.82} \lmk \frac{f_{a}}{10^{12}\,{\rm GeV}}
              \rmk^{0.82} \non \\
         && \times \lmk \frac{\Lambda_{\rm QCD}}{200 {\rm MeV}} 
               \rmk^{-0.65} \lmk \frac{\CN_{\rm QCD}}{60} \rmk^{-0.41}
\:,
\eea

\noindent
with $m_{a}$ being the axion mass at zero temperature, $\Lambda_{\rm
  QCD}$ the energy scale of the QCD phase transition, and $\CN_{\rm
  QCD}$ an effective number of massless degrees of freedom at that
time. These domain-walls bounded by strings also decay by emitting
axions if the QCD anomaly factor is unity \cite{VA}. (Otherwise, these
domain walls would rapidly over-dominate the universe.)

As is seen from Eq.(\ref{eqn:number}), the dominant contribution to the
present axion density is those radiated just before wall domination.
The present density parameter, $\Omega_{\rm s}$,
 of relic axions due to emission from
strings is given by

\beq
  \Omega_{\rm s} \sim 2.7 \lmk \frac{\xi}{\epsilon} \rmk h^{-2} \Delta 
                 \lmk \frac{f_{a}}{10^{12} {\rm GeV}} \rmk^{1.18} \:,
\eeq

\noindent
where $h$ is the Hubble constant in units of 100\,km/sec/Mpc.

With $\xi \sim (1.00 \pm 0.08)$ and $\epsilon^{-1} \sim (0.25\pm0.18)$,
$\Omega_{\rm s}$ is given by

\beq
  \Omega_{\rm s} \sim ( 0.68 \pm 0.46 )\,h^{-2} \Delta 
                 \lmk \frac{f_{a}}{10^{12} {\rm GeV}} \rmk^{1.18} \:.
\eeq

\noindent
The condition that $\Omega_{\rm s} \ltilde 1.0$ constrains the breaking
scale of PQ-symmetry $f_{a}$ as

\beq
  f_{a} \ltilde (1.39\pm0.79)\,h^{1.7} 
                      \times 10^{12}\: {\rm GeV} \:, 
\eeq

\noindent
which reads $f_{a} \ltilde (0.19 - 1.5)\,\times 10^{12}$ GeV for $h =
0.5 - 0.8$. Our constraint lies in between those found in
\cite{sharp} and in \cite{flat}, because the
spectral shape agrees with the former but $\xi = 1$ is substantially
smaller than the case of local strings, $\xi = \CO(10)$, which has been
adopted in \cite{sharp}.

There are two additional contributions to the present energy density
of axions besides that from strings considered above. One is that from
decay of a domain-wall connected by strings \cite{domain} and
estimated as

\beq
   \Omega_{{\rm w}} = \gamma h^{-2} 
                 \lmk \frac{f_{a}}{10^{12} {\rm GeV}} \rmk^{1.18} \:, 
\eeq

\noindent
from which we obtain $f_{a} \ltilde \gamma^{0.85}\,h^{1.7} \times
10^{12}$ GeV with $\gamma$ a factor of the order unity. Note that
$\gamma$ also becomes smaller of the order 10 than the previous
estimation. The other comes from the coherent oscillation of the axion
field \cite{zero2} and is estimated as

\beq
   \Omega_{{\rm co}} = 1.5 \times 10^{\pm 0.4} 
                 \lmk \frac{f_{a}}{10^{12} {\rm GeV}} \rmk^{1.18} \:,
\eeq 

\noindent
which leads to $f_{a} \ltilde (0.33 - 1.5) \times 10^{12}$ GeV.  These
contributions may be comparable with or larger than axions radiated
from axionic strings. Taking all these into account, it is safe to say
that $f_{a} \ltilde 10^{11 - 12}$ GeV.

In summary we have solved equations of motion of a complex scalar field
to clarify cosmological evolution of axionic strings.  As a result we
have confirmed these global strings relax into the scaling solution but
with a number density much smaller than local strings would occupy.  
We have also calculated the energy spectrum of axions generated from
these axionic strings, which agreed with the result of Davis and
Shellard \cite{sharp}.  The constraint on $f_a$  turned out to be in
between those obtained in \cite{sharp} and in
\cite{flat}.

MY is grateful to Professor K. Sato for his encouragement.  This work
was partially supported by the Japanese Grant-in-Aid for Scientific
Research from the Monbusho, Nos.\ 10-04558~(MY), 10640250~(MK),
09740334~(JY) and ``Priority Area: Supersymmetry and Unified Theory of
Elementary Particles(\#707)''(MK \& JY).

\def\NPB#1#2#3{Nucl. Phys. {\bf B#1}, #2 (19#3)}
\def\PLB#1#2#3{Phys. Lett. {\bf B#1}, #2 (19#3)}
\def\PLBold#1#2#3{Phys. Lett. {\bf#1B}, #2 (19#3)}
\def\PRD#1#2#3{Phys. Rev. {\bf D#1}, #2 (19#3)}
\def\PRL#1#2#3{Phys. Rev. Lett. {\bf#1}, #2 (19#3)}
\def\PRT#1#2#3{Phys. Rep. {\bf#1}, #2 (19#3)}
\def\ARAA#1#2#3{Ann. Rev. Astron. Astrophys. {\bf#1}, #2 (19#3)}
\def\ARNP#1#2#3{Ann. Rev. Nucl. Part. Sci. {\bf#1}, #2 (19#3)}
\def\MPL#1#2#3{Mod. Phys. Lett. {\bf #1}, #2 (19#3)}
\def\PZC#1#2#3{Zeit. f\"ur Physik {\bf C#1}, #2 (19#3)}
\def\APJ#1#2#3{Ap. J. {\bf #1}, #2 (19#3)}
\def\AP#1#2#3{{Ann. Phys. } {\bf #1}, #2 (19#3)}
\def\RMP#1#2#3{{Rev. Mod. Phys. } {\bf #1}, #2 (19#3)}
\def\CMP#1#2#3{{Comm. Math. Phys. } {\bf #1}, #2 (19#3)}
\def\PTP#1#2#3{{Prog. Theor. Phys. } {\bf #1}, #2 (19#3)}

\end{document}